%% file: COMPASS_DY_TSAs_2023_arXiv.tex
\documentclass[a4paper,manyauthors,nocleardouble,COMPASS]{cernphprep}

\RequirePackage[T1]{fontenc}

\RequirePackage{graphicx}
\RequirePackage{newtxtext}
\RequirePackage{newtxmath}
\RequirePackage{flushend}
\RequirePackage[numbers,sort&compress]{natbib}

\usepackage{etex}
\usepackage{amsmath,amssymb,epsfig,fleqn,url,color,here,graphics,graphicx,rotating,multirow,wrapfig}
\usepackage{pbox}
\usepackage{dcolumn}
\usepackage{bigstrut}
\usepackage{soul,xcolor}
\usepackage{lineno, xcolor}

\usepackage[section]{placeins}
\usepackage{orcidlink}
\usepackage{tikz}
\usepackage{hyperref}
\hypersetup{
    colorlinks,
    citecolor=blue,
    filecolor=blue,
    linkcolor=blue,
    urlcolor=blue
}

\newcommand{\lf}{\left}
\newcommand{\rg}{\right}

\newcommand{\phiCS}{\varphi _{CS}}
\newcommand{\phiS}{\varphi _S}
\newcommand{\thCS}{\theta _{CS}}
\newcommand{\gvc}{GeV/$c$}

\newcommand{\gvcw}{GeV/$c^2$}

\newcommand{\AU}{A_{U}^{1}}

\newcommand{\AS}{A_{T}^{\sin {\phiS}}}
\newcommand{\ATm}{A_T^{\sin \lf( {{\phiCS} - {\phiS}} \rg)}}
\newcommand{\ATp}{A_T^{\sin \lf( {{\phiCS} + {\phiS}} \rg)}}
\newcommand{\ATtm}{A_T^{\sin \lf( {{2\phiCS} - {\phiS}} \rg)}}
\newcommand{\ATtp}{A_T^{\sin \lf( {{2\phiCS} + {\phiS}} \rg)}}

\newcommand{\mAS}{\sin {\phiS}}
\newcommand{\mATm}{\sin \lf( {{\phiCS} - {\phiS}} \rg)}
\newcommand{\mATp}{\sin \lf( {{\phiCS} + {\phiS}} \rg)}
\newcommand{\mATtm}{\sin \lf( {{2\phiCS} - {\phiS}} \rg)}
\newcommand{\mATtp}{\sin \lf( {{2\phiCS} + {\phiS}} \rg)}

\setstcolor{red}
\begin{document}
\begin{titlepage}
\PHnumber{2023-XXX}
\PHdate{\today}
\title{Final COMPASS results on the transverse-spin-dependent \\azimuthal asymmetries in the pion-induced Drell-Yan process}

\Collaboration{The COMPASS Collaboration}
\ShortAuthor{The COMPASS Collaboration}

\begin{abstract}

The COMPASS Collaboration performed measurements of the Drell-Yan process in 2015 and 2018 using a 190 \gvc\, $\pi^{-}$ beam impinging on a transversely polarised ammonia target. Combining the data of both years, we present final results on the amplitudes of the five azimuthal modulations in the dimuon production cross section.
%
Three of these transverse-spin-dependent azimuthal asymmetries (TSAs) probe the nucleon leading-twist Sivers, transversity, and pretzelosity transverse-momentum dependent (TMD) parton distribution functions (PDFs). The other two are induced by subleading effects.
These TSAs provide unique new inputs for the study of the nucleon TMD PDFs and their universality properties. In particular, the Sivers TSA observed in this measurement is consistent with the fundamental QCD prediction of a sign change of naive time-reversal-odd TMD PDFs when comparing the Drell-Yan process with semi-inclusive measurements of deep inelastic scattering.
Also, within the context of model predictions, the observed transversity TSA is consistent with the expectation of a sign change for the Boer-Mulders function.
\end{abstract}

\vfill
\Submitted{(to be submitted to Phys. Rev. Letters)}
\end{titlepage}
{
\pagestyle{empty}
\clearpage
}
\clearpage

\setcounter{page}{1}
%
%
%
After decades of extensive theoretical studies and experimental efforts, enormous progress has been made in the study of the internal structure of nucleons. However, a full understanding of the nucleon structure in terms of quarks and gluons remains an open challenge.
%
%
%
When exploring the three-dimensional parton structure of hadrons in momentum space, two types of measurements are of particular importance: firstly, the semi-inclusive measurements of hadron production in deep-inelastic lepton-nucleon scattering, $\ell \,N \rightarrow \ell^\prime \,h \, X$, hereafter referred to as SIDIS; secondly, the Drell-Yan process, \textit{i.e.} quark-antiquark annihilation into a pair of oppositely charged leptons (dileptons) in hadron-nucleon collisions, $h \, N \rightarrow \ell\,\bar{\ell}\, X$, hereafter referred to as DY.
%

The cross section can, in both cases, be factorised into convolutions of perturbatively calculable hard-scattering parton cross sections and nonperturbative functions. These are the parton distribution functions (PDFs) describing the distribution of quarks in the target nucleon and either the functions describing the fragmentation of a quark into the observed hadron (in SIDIS) or the quark PDFs of the incoming hadron (in DY)~\cite{Collins:1984kg,Collins:1981uk,Ji:2004wu,Ji:2006br,Ji:2006vf,Collins:2011zzd,Aybat:2011zv,Ma:2013aca,Collins:2016hqq}.

The three-dimensional picture of hadrons involves both the longitudinal and the intrinsic transverse motion of partons inside (un)polarised hadrons, as well as the spin degrees of freedom. Within the leading-twist (twist-2) approximation of pQCD, there exist eight transverse-momentum-dependent (TMD) PDFs of the nucleon describing the distributions of longitudinal and transverse momenta of partons and their correlations with nucleon and quark spins.

The TMD PDFs are expected to be universal and process independent for the set of processes, for which standard TMD factorisation theorems exist (e.g. SIDIS, DY)~\cite{Collins:2011zzd}.
However, within the TMD framework of QCD, the two naively time-reversal-odd TMD PDFs $f_{1T}^\perp$ and $h_{1}^\perp$, \textit{i.e.} the quark Sivers~\cite{Sivers:1989cc} and Boer-Mulders functions, are predicted to have the same magnitudes but opposite signs when comparing DY and SIDIS~\cite{Collins:2002kn, Brodsky:2002cx, Brodsky:2002rv}. The experimental test of this fundamental prediction is a major challenge in hadron physics.

The functions $f_{1T}^\perp$, $h_{1}^\perp$ and other TMD PDFs are accessed via measurements of specific azimuthal asymmetries in SIDIS and DY (for recent reviews see \textit{e.g.} Refs.~\cite{Anselmino:2020vlp,Avakian:2019drf}).
%
%
The COMPASS experiment at CERN~\cite{COMPASS:2007rjf, COMPASS:2010shj} has the unique capability to explore the transverse-spin structure of the nucleon in a kinematic region that is similar for the two alternative experimental approaches. 
This mitigates the uncertainties of scale-dependence (the TMD evolution~\cite{Collins:2011zzd,Aybat:2011ta,Anselmino:2012aa,Echevarria:2014xaa,Scimemi:2019cmh}) in the comparison of the TMD PDFs extracted from these two measurements, while studying their process (in)dependence.

In 2017, COMPASS reported the results of the first ever DY measurement with a polarised target~\cite{COMPASS:2017jbv}.
%
A first measurement of the Sivers effect in $W^{\pm}$ and $Z^{0}$-boson production in collisions of transversely polarised protons at RHIC was reported in 2016 by the STAR collaboration~\cite{Adamczyk:2015gyk}. Both measurements showed some evidence for the sign-change property of the Sivers function.
%
%
New DY data with transversely polarised protons were collected by COMPASS in 2018. In this Letter, we report results from the combined analysis of the COMPASS DY data collected in 2015 and 2018. Both data sets are similar in size.
Following the conventions of Refs.~\cite{Arnold:2008kf,COMPASS:2010shj,COMPASS:2017jbv}, the general expression for the spin-dependent part of the cross section of pion-induced DY lepton-pair production off a transversely polarised nucleon can be written as follows:
%
\begin{eqnarray}\label{eq:DY_xsecLO}\nonumber
   \hspace*{-1.0cm}\frac{d\sigma}{dq^{4}d\Omega}
    &\propto& \lf({F^{1}_{U}}+\,{F^{2}_{U}}\rg)\lf(1 +\, \AU{{\cos}^2}\thCS \rg)\\ \nonumber
   &&\hspace*{-1.0cm}   \times\, \bigg\{ 1 + {S_T}\Big[ D_{1}\AS\mAS
   +\, D_{2}\Big(\ATtp\mATtp
   +\, \ATtm\mATtm \Big) \\ \nonumber
   &&\hspace*{0.5cm}+\, D_{3}
        \Big(\ATp\mATp +\, \ATm\mATm \Big)
   \Big] \bigg\}.
\end{eqnarray}
%
%
Here, $q$ is the four-momentum of the exchanged virtual photon, $F^{1}_{U}$, $F^{2}_{U}$ are polarisation- and azimuth-independent structure functions, and the polar asymmetry $\AU$ (often referred to as $\lambda$) is given as $\AU$=$\lf({F^{1}_{U}}-{F^{2}_{U}}\rg)/\lf({F^{1}_{U}}+{F^{2}_{U}}\rg)$. The subscript ($U$)$T$ denotes (in)dependence on the transverse polarisation of the target. In analogy to SIDIS, the virtual-photon depolarisation factors are given as $D_{1}$=$(1+{{\cos}^2}\thCS)/\lf(1 + \lambda{{\cos }^2}\thCS \rg)$, $D_{2}$=${{\sin}^2}\thCS/\lf(1 + \lambda{{\cos }^2}\thCS \rg)$ and $D_{3}$=${\sin}2\thCS/\lf(1 + \lambda{{\cos }^2}\thCS \rg)$. At leading order of pQCD, within the twist-2 approximation, $F^{2}_{U}=0$ and $\lambda=1$. 
%
The angle $\phiS$ defined in the target rest frame is the relative azimuthal angle between the transverse component of the virtual-photon momentum, $q_T$, and the direction of the nucleon transverse polarisation $S_{T}$ (see Ref.~\cite{COMPASS:2017jbv}). The azimuthal angle $\phiCS$ and the polar angle $\thCS$ of the lepton, as well as its solid angle $\Omega$, are defined in the Collins-Soper frame following Refs.~\cite{Arnold:2008kf,COMPASS:2017jbv}.

In Eq.~(\ref{eq:DY_xsecLO}), the transverse-spin-dependent asymmetries $A_{T}^{w}$ (hereafter referred to as TSAs) are the amplitudes of the azimuthal modulations $w=w(\phiS,\phiCS)$, divided by the polarisation- and azimuth-independent part of the DY cross section and the corresponding depolarisation factor.
The cross section comprises five TSAs. Three of them can be described by contributions from twist-2 TMD PDFs, while the other two arise due to higher-twist PDFs related to quark-gluon correlations, which induce a suppression by a factor $Q^{-1}$.
%
%
The three DY twist-2 TSAs, $A_T^{\sin\phiS}$, $A_T^{\sin(2\phiCS-\phiS)}$ and $A_T^{\sin(2\phiCS+\phiS)}$ are related to the nucleon Sivers ($f_{1T}^\perp$), transversity ($h_1$) and pretzelosity ($h_{1T}^\perp$) TMD PDFs, respectively~\cite{Bacchetta:2006tn,Arnold:2008kf}.
In the Sivers TSA, the nucleon TMD PDFs are convoluted with the spin-independent pion TMD PDFs $f_{1,\pi}$, while for the other two TSAs the convolution involves the pion Boer-Mulders TMD PDFs $h_{1,\pi}^{\perp}$.
For convenience, these TSAs are hereafter called Sivers TSA, transversity TSA and pretzelosity TSA.

In the case of unpolarised-hadron production in SIDIS of leptons off transversely polarised nucleons, the three aforementioned nucleon TMD PDFs induce analogous twist-2 TSAs
~\cite{Kotzinian:1994dv,Bacchetta:2006tn,Arnold:2008kf}. Experimentally, these TSAs were investigated by HERMES using a proton target~\cite{Airapetian:2009ae,HERMES:2020ifk} and by COMPASS using both proton and deuteron targets~\cite{Alekseev:2008aa,Airapetian:2009ae,Adolph:2012sp,Adolph:2014fjw,Adolph:2014zba,Adolph:2016dvl,Parsamyan:2013ug}.
%
%
%
Non-zero quark Sivers and transversity TMD PDFs were extracted from SIDIS measurements, using both collinear~\cite{Anselmino:2005ea, Anselmino:2008sga,Anselmino:2013vqa,Radici:2018iag} and TMD evolution approaches~\cite{Aybat:2011ta,Anselmino:2012aa,Anselmino:2016uie,Echevarria:2014xaa,Sun:2013hua,Bastami:2018xqd,Bastami:2020asv,Bacchetta:2020gko,Cammarota:2020qcw}.
%



%
The dimuon production data were collected by the COMPASS experiment in 2015 and in 2018 using the 190 \gvc\; secondary $\pi^{-}$ beam with an average intensity of $0.7\times10^{8}$ s$^{-1}$, delivered from the M2 beamline in the North Area of the Super Proton Synchrotron (SPS) complex at CERN~\cite{Doble:2017syb}.
Beam particles were scattered off a set of consecutive cylindrical targets, mounted coaxially along the beam axis, which is chosen as z-axis of the spectrometer.
%
The polarised proton (NH$_3$) target consisted of two cylindrical cells, each 55 cm long and 4~cm in diameter~\cite{Andrieux:2022zmk}. The two cells were polarised vertically (transverse to the beam axis) in opposite directions, allowing data to be taken with up and down spin orientations simultaneously. In order to compensate for the differences in the dimuon acceptance of the two cells, the polarisation of the target was periodically reversed. The reversals were performed nearly every two weeks to minimize acceptance variations over time. The target transverse polarisation was preserved using a 0.6~T dipole magnetic field with a relaxation time of about 1000 hours~\cite{Andrieux:2022zmk}.  The magnitude of the average proton polarisation during the 2015 and 2018 measurements was $\langle P_T\rangle\approx0.7$. The resolution of the reconstructed interaction vertex position along the z-axis was estimated to be of order of 10 cm for the DY events produced in the polarised target region. The cells were separated by a 20~cm gap to minimize migration of events from one cell to the other. The
dilution factor, accounting for the fraction of polarizable nucleons in the target and the migration of reconstructed events from one target cell to the other, was calculated to be $\langle f\rangle\approx 0.18$.


A 240 cm long hadron absorber made of aluminium-oxide with a cylindrical tungsten core of 5 cm in diameter was placed 135 cm downstream of the polarised target.
%

The COMPASS spectrometer configuration used during the DY measurements was essentially the same as during SIDIS measurements~\cite{COMPASS:2007rjf,Alekseev:2008aa,Adolph:2016dvl}.
The hadrons produced in pion-nucleon interactions in the target region were mostly stopped by the absorber.
Charged particles were detected by the system of tracking detectors in the two-stage spectrometer.
The COMPASS muon identification systems, consisting of a set of large-area trackers and hadron absorbers, allowed the selection of muon tracks.
%
The triggering of dimuon events required the hit pattern of several hodoscope planes to be consistent with two muon candidates originating from the target region. These hodoscope systems covered a wide acceptance in muon polar angle $\theta_{\mu}$ (8~mrad$ < \theta_{\mu} < 160$~mrad).


%

The physics data taking in 2015 (2018) was performed in nine (eight) periods, each consisting of two consecutive, about week-long sub-periods with opposite target polarisations.
For the analysis presented in this Letter, both 2015 and 2018 data were iteratively reprocessed, improving detector calibrations and alignment, and optimizing the reconstruction settings.

The data collected in each given (sub-)period were analysed independently for possible instabilities of kinematic and azimuthal distributions, which could be due to unnoticed detector or trigger problems.
Dimuon event candidates are selected requiring reconstructed tracks of an incoming pion and at least two oppositely charged outgoing muons associated to a common production vertex.
%
%
Production vertices are required to be within the fiducial volumes of the polarised-target cells.
A set of selection criteria was applied to ensure the quality of the reconstructed tracks, the reliability of the muon identification and to verify that the topology of the dimuon events is consistent with the registered trigger patterns.

%

The dimuon transverse momentum $q_T$ is required to be above $0.4$\,\gvc~to ensure sufficient resolution of the azimuthal angles $\phi_{CS}$ and $\phi_S$.
%
%
%
%
In order to reduce background from two-muon events that are not produced via the DY process, the dimuon mass range was chosen as 4.0~\gvcw\,$ < M_{\mu\mu} < $\,9.0~\gvcw.
%
%
This range was enlarged compared to our previous publication~\cite{COMPASS:2017jbv}, where stricter requirements on the invariant mass range were applied: 4.3~\gvcw\,$ < M_{\mu\mu} < $\,8.5~\gvcw. At lower masses, the background contamination consists of contributions from $\psi^\prime$, $J/\psi$, semi-muonic open-charm decays and combinatorial background. Choosing the upper limit at 9.0~\gvcw\, minimises the contribution of $\Upsilon$-resonances.
Based on Monte-Carlo studies (using the PYTHIA-8 generator and the GEANT-4 based COMPASS setup simulation tool), it was observed that the background contribution depends dominantly on the mass. In the first mass bin (4.0~\gvcw\,$ < M_{\mu\mu} < $\,4.36~\gvcw), the estimated background amounts to about 30\%, with the largest contribution coming from the $\psi^\prime$ tail. It rapidly drops to 6\% in the next mass bin (4.36~\gvcw\,$ < M_{\mu\mu} < $\,5.12~\gvcw). 
Over all the enlarged mass range the background was estimated to be about 10\%, while it was below 5\% using the previous selection~\cite{COMPASS:2017jbv}.
%
Recent COMPASS studies indicate that in the $\psi^\prime$ and $J/\psi$ regions, and between them, the asymmetries are small and compatible with zero within 0.5\%-2\% statistical precision.
Hence, this background represents only a dilution to the DY TSAs. The appropriate weighting factors were evaluated on an event-by-event basis as a function of $M_{\mu\mu}$ and included in the overall dilution factor.
%
%
%
%
%
%
After all selections, about $102000$ dimuons events remained for analysis ($50000$ in 2015 and $52000$ in 2018).

The Bjorken scaling variables related to the beam pion, $x_{\pi}$, and the target nucleon, $x_{N}$, have the following average values: $\langle x_{N} \rangle=0.16$, $\langle x_{\pi} \rangle=0.48$. Hence, the kinematic domain explored by the COMPASS measurement probes mainly the valence quark region, where the expected dominant TMD PDF contributions come from the $u$ quarks of the nucleon and the $\bar{u}$ quark of the incoming $\pi^{-}$. The average values for dimuon transverse momentum ($\langle q_{T} \rangle=1.2$~\gvc) and invariant mass ($\langle M_{\mu\mu} \rangle=5.1$~\gvcw) satisfy the requirements imposed by the factorisation theorems~\cite{Bastami:2020asv}.


For each data-taking year separately, all five TSAs present in the cross section (see Eq.~\ref{eq:DY_xsecLO}) are extracted period by period and then averaged. For each TSA, the 2015 and 2018 results are then combined by calculating the weighted average in each kinematic bin, taking into account the quadratically added statistical and systematic uncertainties.
%
%
%
The extraction of the asymmetries is performed using an extended unbinned maximum likelihood estimator, where all five modulations are fitted simultaneously using dimuon events produced in each target cell for the two directions of the target polarisation.
The estimator is based on the method developed for the COMPASS SIDIS TSA analyses~\cite{Adolph:2012nw}. In this approach, flux and acceptance-dependent systematic uncertainties are minimised.
The TSAs are evaluated in one-dimensional kinematic bins as a function of $x_{N}$, $x_{\pi}$, dimuon Feynman variable $x_{F}$, $q_{T}$ or $M_{\mu\mu}$, integrating over the entire accepted range of all other variables.


In order to evaluate the TSAs, the amplitudes of the modulations are corrected for the depolarisation factors and for the effective proton polarisation $f\cdot\langle P_T\rangle$.
The depolarisation factors and the dilution factor are applied as weights on an event-by-event basis. The depolarisation factors are evaluated using the approximation $\lambda=1$. Known deviations from this assumption with $\lambda$ ranging between 0.5 and 1~\cite{Guanziroli:1987rp,Conway:1989fs,Lambertsen:2016wgj} decrease the normalisation by at most $5\%$.

The largest systematic uncertainties for the TSAs are attributed to residual variations of the experimental conditions. Such instabilities may result in changes of the spectrometer acceptance, which may not be entirely cancelled when combining the data in a given period.
The corresponding systematic effects are quantified by evaluating various types of false asymmetries (for details see Refs.~\cite{Adolph:2012sp,Adolph:2012sn}) and by checking the stability of the results over the periods.
Thorough studies performed separately for the two data-taking years revealed somewhat larger systematic effects and instabilities for 2018 compared to 2015.
The systematic point-to-point uncertainties associated with the TSAs were estimated to be between 0.7 to 0.8 times the corresponding statistical uncertainties in 2015 and between 1.0 to 1.2 in 2018.
For the two years, the normalisation uncertainties associated with target polarisation and overall dilution factor are $5\%$ and $12\%$, respectively.
%
%
%
\begin{figure}[h!]
\centering
\includegraphics[width=0.8\textwidth]{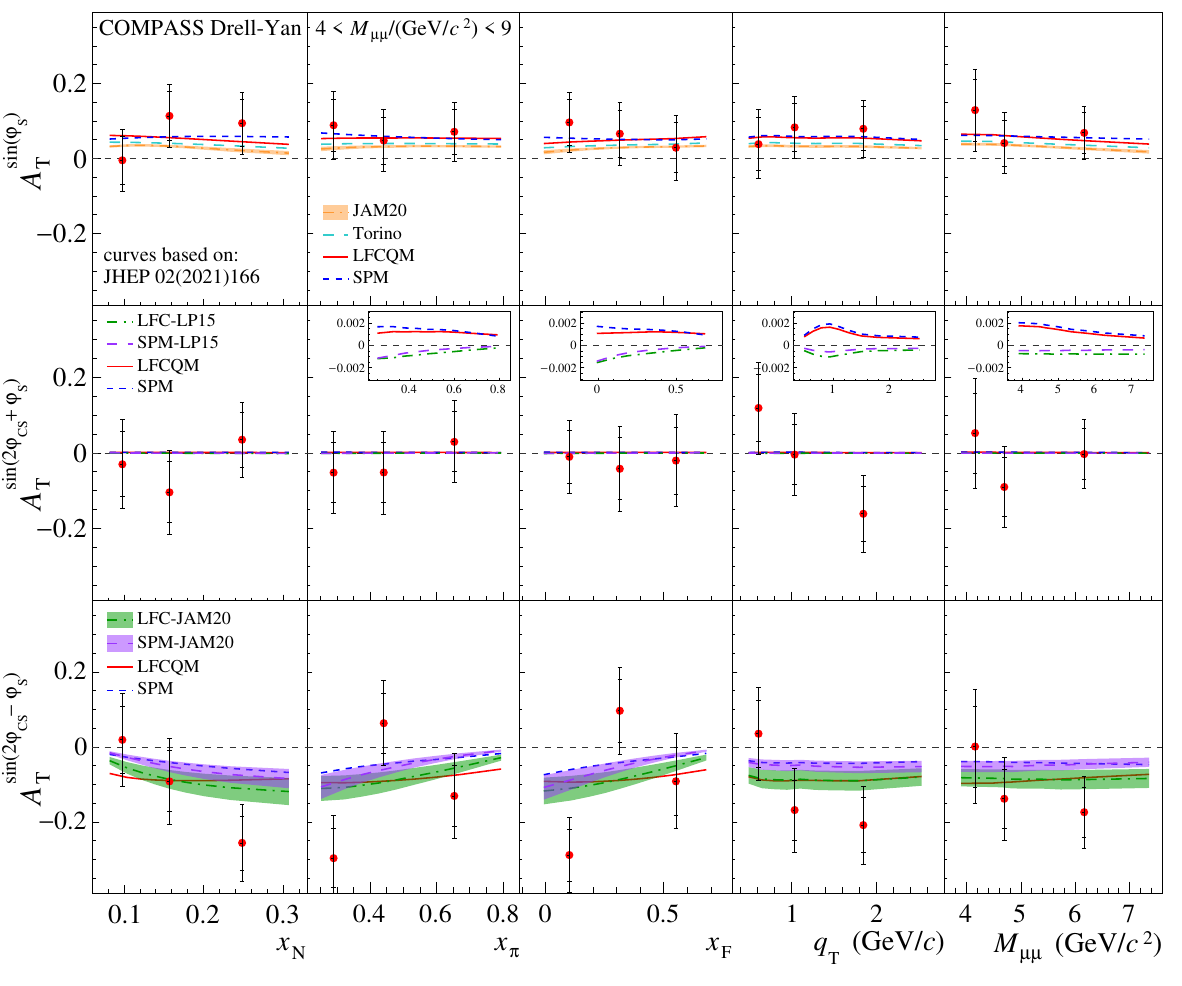}
\caption{Kinematic dependences of the Sivers, pretzelosity, and transversity TSAs (top to bottom).
Inner (outer) error bars represent statistical (total experimental) uncertainties. For theoretical predictions see text.}
\label{fig:TSAs}
\end{figure}

In Fig.~\ref{fig:TSAs}, the combined 2015 and 2018 COMPASS results obtained for the three twist-2 TSAs $A_T^{\sin\phiS}$, $A_T^{\sin(2\phiCS+\phiS)}$ and $A_T^{\sin(2\phiCS-\phiS)}$ are shown as a function of the variables $x_{N}$, $x_{\pi}$, $x_{F}$, $q_{T}$ and $M_{\mu\mu}$.
%
Compared to the previous analysis of only the 2015 data~\cite{COMPASS:2017jbv}, adding the 2018 data and enlarging the dimuon mass range increased the statistical precision of the measurement by a factor of 1.5.
The presented TSAs are compared with recent theoretical predictions, which are based on calculations performed in Ref.~\cite{Bastami:2020asv}.
These predictions are obtained by using for each bin the appropriate average kinematic values given by the event population.
For each TSA, four different calculations based on two different approaches are presented.
The first approach is solely based on model predictions for pion and proton TMD PDFs using the light-front constituent quark model (LFCQM)~\cite{Lorce:2016ugb} and the spectator model (SPM)~\cite{Jakob:1997wg,Lu:2004hu,Gamberg:2007wm}.
The second is a ``hybrid'' approach, in which model inputs are restricted to the usage of LFCQM and SPM for the pion Boer-Mulders function, while the non-perturbative inputs for the proton TMD PDFs are taken from available parameterisations extracted from experimental data (``Torino'' fit~\cite{Anselmino:2013vqa}, ``JAM20'' global fit~\cite{Cammarota:2020qcw} and ``LP15'' fit~\cite{Lefky:2014eia}). The  MSTW extraction~\cite{Anselmino:2013lza} was used for the collinear proton PDF $f_{1,p}$, while for the collinear pion PDF $f_{1,\pi}$ the SMRS~\cite{Sutton:1991ay} fits were used.
In these predictions, the TMD evolution is implemented at next-to-leading logarithmic precision for all twist-2 TSAs.
%
The model calculations were performed using the sign-change hypothesis for both the nucleon Sivers and Boer-Mulders TMD PDFs~\cite{Peng:2014pha,Bastami:2020asv}.

\begin{figure}[h!]
\centering
\includegraphics[width=0.49\textwidth]{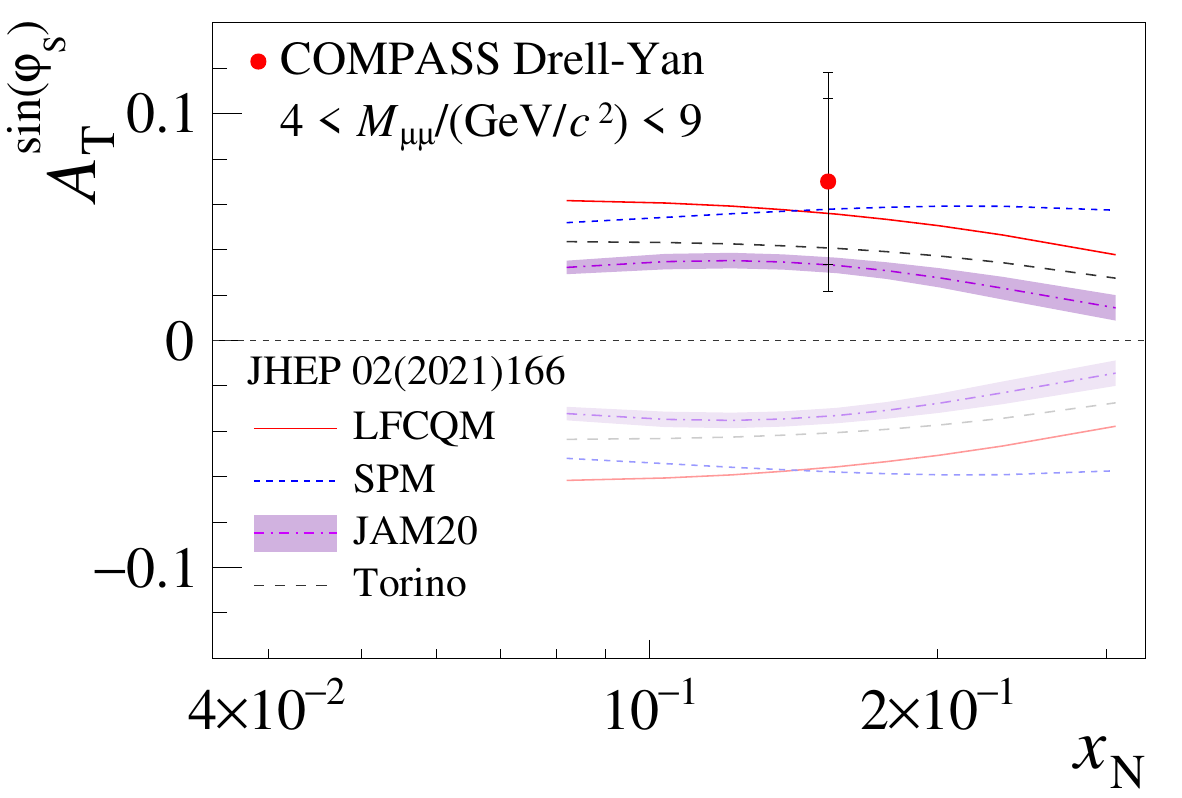}
\includegraphics[width=0.49\textwidth]{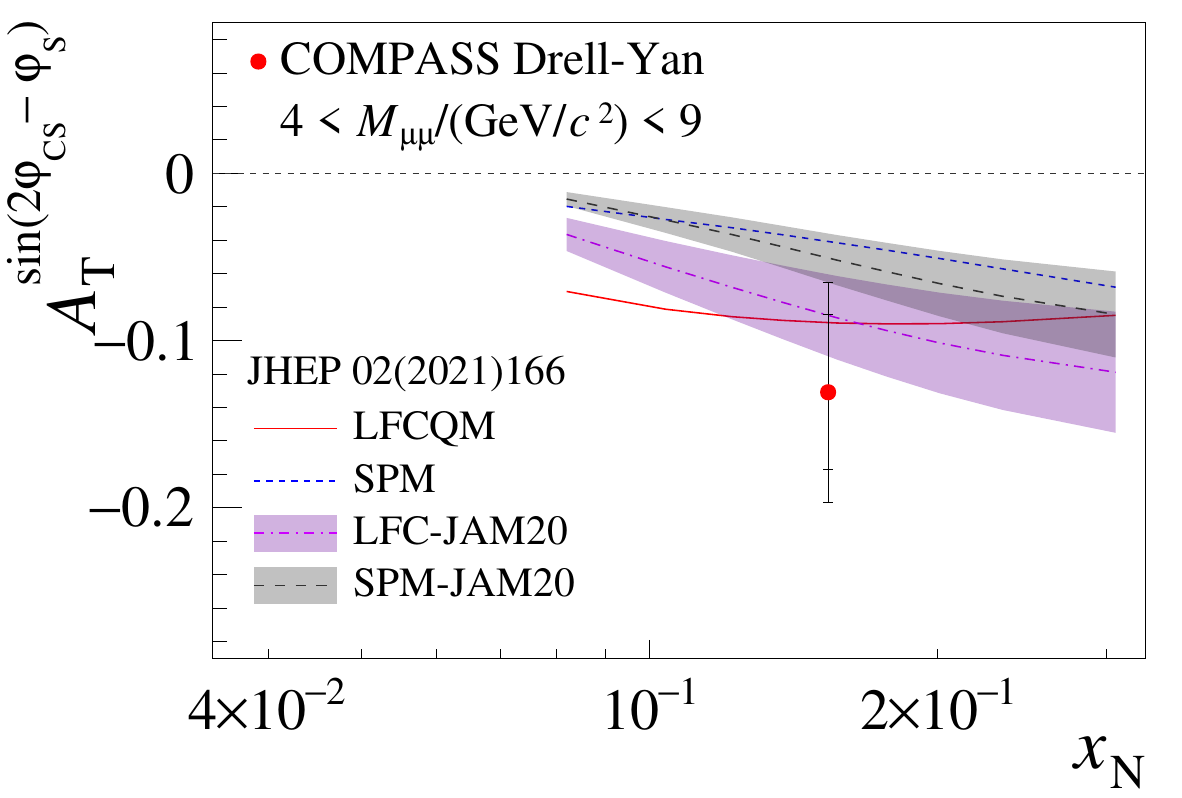}
\caption{Left panel: Measured average Sivers TSA and theoretical  predictions from different models from Ref.~\cite{Bastami:2020asv}.
The dark-shaded (light-shaded) predictions are evaluated with (without) the sign-change hypothesis. Right panel: Measured average transversity TSA and theoretical  predictions from different models from Ref.~\cite{Bastami:2020asv}. Otherwise as in Fig.~\ref{fig:TSAs}}.
\label{fig:SivTr_theor}
\end{figure}

The Sivers TSA $A_T^{\sin\phiS}$ is predicted to be positive in the entire kinematic range~\cite{Bastami:2020asv},
which is in agreement with the COMPASS data points shown in Fig.~\ref{fig:TSAs}.
The average Sivers TSA, $\langle A_T^{\sin\phiS}\rangle=0.070\pm0.037(stat.)\pm0.031(sys.)$, is found to be above zero at about 1.5 standard deviations of the total uncertainty.
In the left panel of Fig.~\ref{fig:SivTr_theor}, the Sivers TSA is shown together with model predictions~\cite{Bastami:2020asv} evaluated with and without the sign-change hypothesis, shown as dark-shaded curves in the top and light-shaded curves in the bottom of the figure, respectively.
Using the band of the presented model predictions, the COMPASS measurement is found to agree with the sign-change hypothesis within less than one standard deviation of its total uncertainty, while being away from the no-sign-change hypothesis by about 2.5 to 3 standard deviations.
In addition, the present results do not support earlier expectations of a large Sivers effect in the DY process at COMPASS kinematics~\cite{COMPASS:2010shj}.

The transversity TSA $A_T^{\sin(2\phiCS-\phiS)}$ is expected to be negative, but larger in absolute value compared to the Sivers TSA~\cite{Sissakian:2010zza,Bastami:2020asv}. 
The average value for the transversity TSA is measured to be below zero with a significance of about two standard deviations, $\langle A_T^{\sin(2\phiCS-\phiS)}\rangle=-0.131\pm0.046(stat.)\pm0.047(sys.)$.
In the right panel of Fig.~\ref{fig:SivTr_theor}, the average transversity TSA is shown together with model calculations~\cite{Bastami:2020asv}. The COMPASS measurement is found to agree in sign and magnitude with the band of available model predictions, which supports the universal nature of the transversity TMD PDFs.
%
%
In the context of Refs.~\cite{Peng:2014pha,Bastami:2020asv} this observation helps to fix the sign of the $\bar{u}$ quark pion Boer-Mulders TMD PDF and consequently also supports the sign change for the nucleon Boer-Mulders TMD PDFs.
%
Altogether, the above discussion supports the general validity of the TMD approach used to evaluate the predictions.

%
%
%
%

%

The pretzelosity TSA $A_T^{\sin(2\phiCS+\phiS)}$ is predicted to be very small, which is explained by the magnitude of the pretzelosity TMD PDFs and kinematic suppression factors~\cite{Bastami:2020asv}.
The measured average value, $\langle A_T^{\sin(2\phiCS+\phiS)}\rangle=-0.027\pm0.046(stat.)\pm0.043(sys.)$, is indeed found to be small and compatible with zero within uncertainties.
For the two higher-twist TSAs, the averaged values $\langle A_T^{\sin(\phiCS-\phiS)}\rangle=0.113\pm0.076(stat.)\pm0.071(sys.)$ and $\langle A_T^{\sin(\phiCS+\phiS)}\rangle=-0.071\pm0.071(stat.)\pm0.064(sys.)$ are consistent with zero within about one standard deviation of the total uncertainty.
%
Compared to the twist-2 TSAs, the statistical uncertainties of the two TSAs related to higher-twist TMD PDFs are notably larger, which is explained by the relative smallness of the depolarisation factor $D_{3}$.
No predictions are available for these twist-3 TSAs.

The full set of numerical values for all TSAs, including correlation coefficients and average kinematic values from this measurement, is available on HepData~\cite{hepdata}.
This includes also the results for the narrower mass range used in our previous publication~\cite{COMPASS:2017jbv}.

The new COMPASS results presented in this Letter supersede the previous ones from our first publication~\cite{COMPASS:2017jbv}. 
They demonstrate the importance and the potential of measuring the DY process with transversely polarised nucleon targets, thereby paving the way for new projects aiming to perform similar studies at CERN and elsewhere~\cite{Aidala:2019pit,Barschel:2020drr,Keller:2022abm}.

%
%

%
%
\section{Acknowledgments}

We gratefully acknowledge the support of the CERN management and staff and the skill and effort of the technicians of our collaborating institutes.
We are grateful to S. Bastami and the Authors of Ref.~\cite{Bastami:2020asv} for providing us with numerical values of their model predictions.


%
\input{COMPASS_DY_TSAs_2023_arXiv.bbl}
\clearpage
\include{Authors_list_2023.12.10_PRL_DY_TSAs}

\end{document}

%% file: COMPASS_DY_TSAs_2023_arXiv.bbl
%

%% file: Authors_list_2023.12.10_PRL_DY_TSAs.tex
\center{\textbf{The COMPASS Collaboration}}

\vspace{10pt}
\begin{flushleft}
G.~D.~Alexeev$^\textrm{{\footnotesize\hyperlink{hl:dubna}{29}}}$\orcidlink{0009-0007-0196-8178},
M.~G.~Alexeev$^\textrm{{\footnotesize\hyperlink{hl:turin_u}{20},\hyperlink{hl:turin_i}{19}}}$\orcidlink{0000-0002-7306-8255},
C.~Alice$^\textrm{{\footnotesize\hyperlink{hl:turin_u}{20},\hyperlink{hl:turin_i}{19}}}$\orcidlink{0000-0001-6297-9857},
A.~Amoroso$^\textrm{{\footnotesize\hyperlink{hl:turin_u}{20},\hyperlink{hl:turin_i}{19}}}$\orcidlink{0000-0002-3095-8610},
V.~Andrieux$^\textrm{{\footnotesize\hyperlink{hl:illinois}{34}}}$\orcidlink{0000-0001-9957-9910},
V.~Anosov$^\textrm{{\footnotesize\hyperlink{hl:dubna}{29}}}$\orcidlink{0009-0003-3595-9561},
K.~Augsten$^\textrm{{\footnotesize\hyperlink{hl:praguectu}{4}}}$\orcidlink{0000-0001-8324-0576},
W.~Augustyniak$^\textrm{{\footnotesize\hyperlink{hl:warsaw}{24}}}$,
C.~D.~R.~Azevedo$^\textrm{{\footnotesize\hyperlink{hl:aveiro}{27}}}$\orcidlink{0000-0002-0012-9918},
B.~Badelek$^\textrm{{\footnotesize\hyperlink{hl:warsawu}{26}}}$\orcidlink{0000-0002-4082-1466}
% ...
\hypertarget{hl:calcutta}{$^\textrm{{\footnotesize 14}}$\footnotesize~Matrivani Institute of Experimental Research \& Education, Calcutta-700 030, India$^\textrm{{\tiny\hyperlink{hl:C}{C}}}$\\}
\hypertarget{hl:telaviv}{$^\textrm{{\footnotesize 15}}$\footnotesize~Tel Aviv University, School of Physics and Astronomy, 69978 Tel Aviv, Israel$^\textrm{{\tiny\hyperlink{hl:D}{D}}}$\\}
\hypertarget{hl:triest_a}{$^\textrm{{\footnotesize 16}}$\footnotesize~Abdus Salam ICTP, 34151 Trieste, Italy\\}
\hypertarget{hl:triest_i}{$^\textrm{{\footnotesize 17}}$\footnotesize~Trieste Section of INFN, 34127 Trieste, Italy\\}
\hypertarget{hl:triest_u}{$^\textrm{{\footnotesize 18}}$\footnotesize~University of Trieste, Dept.\ of Physics, 34127 Trieste, Italy\\}
\hypertarget{hl:turin_i}{$^\textrm{{\footnotesize 19}}$\footnotesize~Torino Section of INFN, 10125 Torino, Italy\\}
\hypertarget{hl:turin_u}{$^\textrm{{\footnotesize 20}}$\footnotesize~University of Torino, Dept.\ of Physics, 10125 Torino, Italy\\}
\hypertarget{hl:miyazaki}{$^\textrm{{\footnotesize 21}}$\footnotesize~University of Miyazaki, Miyazaki 889-2192, Japan$^\textrm{{\tiny\hyperlink{hl:E}{E}}}$\\}
\hypertarget{hl:nagoya}{$^\textrm{{\footnotesize 22}}$\footnotesize~Nagoya University, 464 Nagoya, Japan$^\textrm{{\tiny\hyperlink{hl:E}{E}}}$\\}
\hypertarget{hl:yamagata}{$^\textrm{{\footnotesize 23}}$\footnotesize~Yamagata University, Yamagata 992-8510, Japan$^\textrm{{\tiny\hyperlink{hl:E}{E}}}$\\}
\hypertarget{hl:warsaw}{$^\textrm{{\footnotesize 24}}$\footnotesize~National Centre for Nuclear Research, 02-093 Warsaw, Poland$^\textrm{{\tiny\hyperlink{hl:F}{F}}}$\\}
\hypertarget{hl:warsawtu}{$^\textrm{{\footnotesize 25}}$\footnotesize~Warsaw University of Technology, Institute of Radioelectronics, 00-665 Warsaw, Poland$^\textrm{{\tiny\hyperlink{hl:F}{F}}}$\\}
\hypertarget{hl:warsawu}{$^\textrm{{\footnotesize 26}}$\footnotesize~University of Warsaw, Faculty of Physics, 02-093 Warsaw, Poland$^\textrm{{\tiny\hyperlink{hl:F}{F}}}$\\}
\hypertarget{hl:aveiro}{$^\textrm{{\footnotesize 27}}$\footnotesize~University of Aveiro, I3N, Dept. of Physics, 3810-193 Aveiro, Portugal$^\textrm{{\tiny\hyperlink{hl:G}{G}}}$\\}
\hypertarget{hl:lisbon}{$^\textrm{{\footnotesize 28}}$\footnotesize~LIP, 1649-003 Lisbon, Portugal$^\textrm{{\tiny\hyperlink{hl:G}{G}}}$\\}
\hypertarget{hl:dubna}{$^\textrm{{\footnotesize 29}}$\footnotesize~Affiliated with an international laboratory covered by a cooperation agreement with CERN\\}
\hypertarget{hl:russia}{$^\textrm{{\footnotesize 30}}$\footnotesize~Affiliated with an institute covered by a cooperation agreement with CERN.\\}
\hypertarget{hl:cern}{$^\textrm{{\footnotesize 31}}$\footnotesize~CERN, 1211 Geneva 23, Switzerland\\}
\hypertarget{hl:taipei}{$^\textrm{{\footnotesize 32}}$\footnotesize~Academia Sinica, Institute of Physics, Taipei 11529, Taiwan$^\textrm{{\tiny\hyperlink{hl:H}{H}}}$\\}
\hypertarget{hl:taipeincu}{$^\textrm{{\footnotesize 33}}$\footnotesize~Center for High Energy and High Field Physics and Dept.\ of Physics, National Central University, 300 Zhongda Rd., Zhongli 320317, Taiwan$^\textrm{{\tiny\hyperlink{hl:H}{H}}}$\\}
\hypertarget{hl:illinois}{$^\textrm{{\footnotesize 34}}$\footnotesize~University of Illinois at Urbana-Champaign, Dept.\ of Physics, Urbana, IL 61801-3080, USA$^\textrm{{\tiny\hyperlink{hl:I}{I}}}$\\}

\vspace{10pt}
\hypertarget{hl:*}{$^\textrm{{\footnotesize *}}$\footnotesize~Corresponding author\\}
\hypertarget{hl:a}{$^\textrm{{\footnotesize a}}$\footnotesize~Supported by the European Union’s Horizon 2020 research and innovation programme under grant agreement STRONG–2020 - No 824093\\}
\hypertarget{hl:b}{$^\textrm{{\footnotesize b}}$\footnotesize~Retired from Ludwig-Maximilians-Universit\"at, 80539 M\"unchen, Germany\\}
\hypertarget{hl:b1}{$^\textrm{{\footnotesize b1}}$\footnotesize~Supported by the DFG cluster of excellence `Origin and Structure of the Universe' (www.universe-cluster.de) (Germany)\\}
\hypertarget{hl:c}{$^\textrm{{\footnotesize c}}$\footnotesize~Also at ORIGINS Excellence Cluster, 85748 Garching, Germany\\}
\hypertarget{hl:d}{$^\textrm{{\footnotesize d}}$\footnotesize~Also at Institut f\"ur Theoretische Physik, Universit\"at T\"ubingen, 72076 T\"ubingen, Germany\\}
\hypertarget{hl:e}{$^\textrm{{\footnotesize e}}$\footnotesize~Present address: NISER, Centre for Medical and Radiation Physics, Bubaneswar, India\\}
\hypertarget{hl:f}{$^\textrm{{\footnotesize f}}$\footnotesize~Also at University of Eastern Piedmont, 15100 Alessandria, Italy\\}
\hypertarget{hl:f1}{$^\textrm{{\footnotesize f1}}$\footnotesize~Supported by the Funds for Research 2019-22 of the University of Eastern Piedmont\\}
\hypertarget{hl:g}{$^\textrm{{\footnotesize g}}$\footnotesize~Also at Chubu University, Kasugai, Aichi 487-8501, Japan\\}
\hypertarget{hl:h}{$^\textrm{{\footnotesize h}}$\footnotesize~Also at KEK, 1-1 Oho, Tsukuba, Ibaraki 305-0801, Japan\\}
\hypertarget{hl:i}{$^\textrm{{\footnotesize i}}$\footnotesize~Also at Dept.\ of Physics, Pusan National University, Busan 609-735, Republic of Korea\\}
\hypertarget{hl:i1}{$^\textrm{{\footnotesize i1}}$\footnotesize~Also at Physics Dept., Brookhaven National Laboratory, Upton, NY 11973, USA\\}
\hypertarget{hl:j}{$^\textrm{{\footnotesize j}}$\footnotesize~Also at Fairmont State University, Department of Natural Sciences, 1201 Locust Ave, Fairmont, West Virginia 26554, USA\\}
\hypertarget{hl:k}{$^\textrm{{\footnotesize k}}$\footnotesize~Also at Dept.\ of Physics, National Kaohsiung Normal University, Kaohsiung County 824, Taiwan\\}
\hypertarget{hl:$\dagger$}{$^\textrm{{\footnotesize $\dagger$}}$\footnotesize~Deceased\\}

\vspace{10pt}
\hypertarget{hl:A}{$^\textrm{{\tiny A}}$\footnotesize~Supported by MEYS Grants LM2018104, LM2023040 and LTT17018 and Charles University grants PRIMUS/22/SCI/017 and GAUK60121 (Czech Republic)\\}
\hypertarget{hl:B}{$^\textrm{{\tiny B}}$\footnotesize~Supported by BMBF - Bundesministerium f\"ur Bildung und Forschung (Germany)\\}
\hypertarget{hl:C}{$^\textrm{{\tiny C}}$\footnotesize~Supported by B. Sen fund (India)\\}
\hypertarget{hl:D}{$^\textrm{{\tiny D}}$\footnotesize~Supported by the Israel Academy of Sciences and Humanities (Israel)\\}
\hypertarget{hl:E}{$^\textrm{{\tiny E}}$\footnotesize~Supported by MEXT and JSPS, Grants 18002006, 20540299, 18540281 and 26247032, the Daiko and Yamada Foundations (Japan)\\}
\hypertarget{hl:F}{$^\textrm{{\tiny F}}$\footnotesize~Supported by NCN, Grant 2020/37/B/ST2/01547 (Poland)\\}
\hypertarget{hl:G}{$^\textrm{{\tiny G}}$\footnotesize~Supported by FCT, Grants DOI 10.54499/CERN/FIS-PAR/0022/2019 and DOI 10.54499/CERN/FIS-PAR/0016/2021 (Portugal)\\}
\hypertarget{hl:H}{$^\textrm{{\tiny H}}$\footnotesize~Supported by the Ministry of Science and Technology (Taiwan)\\}
\hypertarget{hl:I}{$^\textrm{{\tiny I}}$\footnotesize~Supported by the National Science Foundation, Grant no. PHY-1506416 (USA)\\}

\end{flushleft}